\documentclass[11pt]{article}

\usepackage{scalerel}
\usepackage{amssymb,amsmath,amsthm,mathrsfs,latexsym,amsxtra,graphicx,appendix,epstopdf,feynmf,subfig,hyperref,setspace,fix-cm,color,cite,hyperref}
\usepackage{verbatim}
\usepackage[a4paper,left=2cm,right=1cm,top=4cm,bottom=4cm,bindingoffset=5mm]{geometry}
\usepackage{pgfplots}
\pgfplotsset{compat=newest}
\usepackage{mathtools}
\usepackage{tensor}
\usepackage{siunitx}
\usepackage{tikz}
\usetikzlibrary{trees}
\usetikzlibrary{decorations.pathmorphing}
\usetikzlibrary{decorations.markings}
\usetikzlibrary{decorations.markings}
\usepackage{float}
\usepackage{rotating}
\usepackage{simplewick}
\usepackage{cancel}
\usepackage{ marvosym }
\usepgfplotslibrary{fillbetween}
\usetikzlibrary{patterns}

\usepackage{mathtools}

\usepackage{setspace}
\usepackage{lipsum}
 \usepackage{multirow}
\usepackage{fullpage}
\usepackage{amsmath}
\usepackage{amssymb}
\usepackage{setspace}
\usepackage{bbm}
\usepackage{dsfont}
\usepackage{graphics}
\usepackage{longtable}
\usepackage[font=footnotesize,labelfont=bf,justification=centerlast,width=.9\textwidth]{caption}
\usepackage{color}
\usepackage{etoolbox}
 \usepackage{multirow}
\usepackage{booktabs}
\usepackage{array}
\usepackage{hyperref}
\usepackage{cite}
\usepackage[normalem]{ulem}
\hypersetup{
    bookmarks=true,%
    colorlinks,%
    citecolor=blue,%
    filecolor=blue,%
    linkcolor=blue,%
    urlcolor=blue
}
\allowdisplaybreaks
\newcommand{\twobytwo}[4]{\left(\begin{array}{cc}#1&#2\\#3&#4\end{array}\right)}
\newcommand\rref[1]{(\ref{#1})}
\newcommand{\be}{\begin{equation}}
\newcommand{\ee}{\end{equation}}
\newcommand{\bes}{\begin{equation*}}
\newcommand{\ees}{\end{equation*}}
\newcommand{\bea}{\begin{eqnarray}}
\newcommand{\eea}{\end{eqnarray}}
\newcommand{\beas}{\begin{eqnarray*}}
\newcommand{\eeas}{\end{eqnarray*}}

\newcommand{\lns}{{l_{\rm NS}}}
\newcommand{\NS}{\text{NS}}
\newcommand{\dinf}{d_\infty^{\text{NS}}(h,\lns)}

\newcommand{\bmat}{\begin{bmatrix}}
\newcommand{\emat}{\end{bmatrix}}

\def\le{\left}
\def\ri{\right}

\newcommand{\CC}{\mathbb{C}}

\newcommand{\ZZ}{\mathbb{Z}}

\newcommand{\N}{\mathcal{N}}

\newcommand{\Tr}{{\rm {Tr}}}

\renewcommand{\H}{\mathbb{H}}
\newcommand{\Z}{\mathbb{Z}}

\newcommand{\ie}{{\it i.e.~}}

\newcommand{\tf}{{\tilde{f}}}

\newcommand{\fns}{f_{\textrm{NS}}}

\begin{document}
\numberwithin{equation}{section}
{
\begin{titlepage}

\hfill { \large CERN-TH-2019-164 }

\begin{center}

\hfill \\
\hfill \\
\vskip 0.75in

{\Large \bf The Holographic Landscape of Symmetric Product Orbifolds}

\vskip 0.4in

{\large Alexandre Belin${}^a$, Alejandra Castro${}^b$, Christoph A.~Keller${}^{c}$, and  Beatrix M\"uhlmann${}^b$}\\
\vskip 0.3in

${}^{a}${\it CERN, Theory Division, 1 Esplanade des Particules, Geneva 23, CH-1211, Switzerland}
\vskip .5mm
${}^{b}${\it Institute for Theoretical Physics, University of Amsterdam,
Science Park 904, Postbus 94485, 1090 GL Amsterdam, The Netherlands} \vskip .5mm

${}^{c}${\it Department of Mathematics, University of Arizona, Tucson, AZ 85721-0089, USA} \vskip .5mm
\texttt{a.belin@cern.ch, a.castro@uva.nl, cakeller@math.arizona.edu, b.muhlmann@uva.nl}

\end{center}

\vskip 0.35in

\begin{center} {\bf ABSTRACT } \end{center}
We investigate the growth of coefficients in the elliptic genus of symmetric product orbifolds at large central charge. We find that this landscape decomposes into two regions. In one region, the growth of the low energy states is Hagedorn, which indicates a stringy dual. In the other, the growth is much slower, and compatible with the spectrum of a supergravity theory on AdS$_3$. 
We provide a simple diagnostic which places any symmetric product orbifold in either region.
We construct a class of elliptic genera with such supergravity-like growth, indicating the possible existence of new realizations of AdS$_3$/CFT$_2$ where the bulk is a semi-classical supergravity theory.   
In such cases, we give exact expressions for the BPS degeneracies, which could be matched with the spectrum of perturbative states in a dual supergravity description.
 \vfill
\noindent \today

\end{titlepage}
}

\newpage

\tableofcontents

\section{Introduction}

The AdS/CFT correspondence provides a non-perturbative, UV-complete definition of quantum gravity in Anti-de Sitter space. In the realizations that are well understood, it relates supergravity theories on backgrounds of the type AdS$_{d+1}\times M_q$ to conformal field theories living in $d$ dimensions. The conformal field theories at play have striking features such as a large number of degrees of freedom and strong coupling, which are often seen as necessary conditions to make CFTs ``holographic." Given the space of all conformal field theories, one would like to understand which CFTs possess the appropriate properties to be described by semi-classical general relativity (or supergravity). This remains one of the most important open questions in holography.

In this work, we will address this question in the context of AdS$_3$/CFT$_2$. The advantage of working in this low-dimensional setting is that its symmetries give us more control while retaining most of the flavor of its higher dimensional counterparts. Some of the universal aspects controlled by symmetries are the following. Two-dimensional conformal field theories are constrained by the infinite-dimensional Virasoro algebra, which as shown several decades ago by Brown and Henneaux \cite{Brown:1986nw}, gives a universal relation between the value of the central charge and the gravitational coupling
\be
c=\frac{3\ell_{\rm AdS}}{2G_N} \,.
\ee
Semi-classical theories of gravity must  therefore correspond to CFTs with a large value of the central charge, \ie CFTs with a large number of degrees of freedom. A CFT$_2$ is also constrained by modular invariance, which in particular implies that at large energies $h$ the  asymptotic behavior of the density of states $\rho(h)$  is completely fixed by symmetry. This behavior is given by the Cardy formula \cite{cardyformula}, i.e. 
\be\label{eq:rho1} 
\rho(h) \approx e^{2\pi \sqrt{\frac{c}{3}2h}}~,\qquad h=\bar{h}\gg 1~, \quad c~{\rm fixed}\,.
\ee
Famously, this matches the entropy of the BTZ black hole \cite{Strominger:1997eq}. While this provides a consistency condition on AdS/CFT, it is somewhat of an expected one since it is guaranteed by manifest symmetries on both sides of the duality. Moreover, the match is insensitive to the CFT: \eqref{eq:rho1} holds independently of the value of the central charge or other dynamical features of the theory.

Moving away from asymptotically large energies $h$, the spectrum of a CFT$_2$ is no longer universal and provides a window into the dynamics of the theory. From the spectrum alone, one can derive necessary conditions for a CFT to behave holographically by matching with gravitational results. A first step was carried out in \cite{Hartman:2014oaa} by demanding that the CFT entropy match that of the BTZ black hole not only at asymptotically large energies, but for any black hole above the Hawking-Page phase transition. Demanding such a matching requires the density of light states to be sufficiently sparse, namely it should satisfy the HKS bound
\be \label{HKS}
\rho(h) \lesssim e^{2\pi h} \,, \qquad h <\frac{c}{12} \,.
\ee
Similar ideas were later carried out for other types of observables in, e.g., \cite{Benjamin:2015hsa,Belin:2016yll,Belin:2017nze,Kraus:2017kyl,Mefford:2017oxy,Anous:2018hjh,Michel:2019vnk}.

While the HKS bound is necessary for a CFT to be holographic, it is not sufficient. In particular, \eqref{HKS} allows for a Hagedorn growth of light states. This growth is typical of a string theory in AdS, and would produce large deviations from general relativity at low energies. In this paper, the precise form of the growth of light states $\rho(h)$ will be the key object of study. Light states are perturbative states dual to particles in the low-energy bulk effective field theory. Their growth teaches us about the nature of the effective field theory that lives in the bulk.\footnote{The spectrum of light particles also enters into the data necessary to compute correlation functions, once we consider loops. This connection was exploited in \cite{Alday:2019qrf} to read off the number of spacetime dimensions from CFT correlators.} We will mostly distinguish two scenarios.
\begin{enumerate}
\item {\bf Hagedorn growth}:  $\qquad \qquad \qquad \rho(h)\sim e^{ c_H h} $. 
\item  {\bf Supergravity-like growth}: $\ \qquad \rho(h)\sim e^{ c_S h^{\alpha}} \,, \quad \alpha <1$.
\end{enumerate}
The spectra described above should be viewed as an asymptotic growth for light states, by which we mean
\be
1 \ll h \ll c \,.
\ee
The two parameters $c_H$ and $c_S$ are left unspecified at this point. The idea behind the supergravity-like growth is that a weakly interacting local quantum field theory living in AdS$_3\times M_{D-3}$ would have a spectrum that grows like
\be\label{eq:sugrarho}
\rho(h) \sim e^{c_S h^\frac{D-1}{D}} \,.
\ee
The main goal of this paper is to construct examples that possess this type of growth. While obtaining theories that exhibit Hagedorn growth (and hence are compatible with the HKS bound) is easy,  only a handful of CFTs are known to comply to a supergravity-like spectrum. Here we will identify and quantify a large class of counting formulas that exhibit the desired supergravity-like growth, and in particular we will give explicit expressions for the density of states.

We note that all our supergravity-like examples will turn out to have
$\alpha=1/2$. The mathematical reason for this is explained in
\cite{Belin:2019jqz}. The physical reason is that we will be computing and index for which certain states cancel out. This  leads to a lower effective
dimension $D$. Why and how this happens in detail is not completely
clear to us, but we note that this is exactly what happens in the
known examples with $D=6$ such as the D1D5 CFT dual to supergravity on AdS$_3\times S^3$ \cite{Benjamin:2016pil}: Restricting to
supergravity chiral primaries, $\alpha$ reduces from $5/6$ to
$3/4$, which is exactly the growth that appears in the so-called
Hodge elliptic genus \cite{Kachru:2016igs}. To obtain the
elliptic genus, which is the object that we investigate in the present
work, we then specialize the Hodge elliptic genus. This specialization
leads to cancellations between BPS states, which further reduces
$\alpha$ to $1/2$. It is not clear to us how this mechanism works
for our more general examples. We hope to understand this in future work.

\subsection{Symmetric product orbifolds: Hagedorn vs supergravity growth}

Our objective should now be clear: we want to scan the space of CFT$_2$ with a large central charge and ask whether the low-lying spectrum is supergravity-like. Even though a CFT$_2$ is highly constrained due to the infinite dimensional Virasoro symmetry, we unfortunately lack a complete classification of all consistent theories. Still, we have certain regions of theory-space that we can explore. One rich region is the space of CFTs  that admit a description in terms of a symmetric product orbifold. This class of theories is constructed as follows: one considers a seed CFT $\mathcal{C}$, takes $N$ copies, and orbifolds by the symmetric group that exchanges the copies
\be
\mathcal{C}_N = \frac{\mathcal{C}^{\otimes N}}{S_N} \,.
\ee
These theories are known to possess a good large $N$ limit \cite{Pakman:2009zz,Haehl:2014yla,Belin:2014fna}, and hence provide an interesting space of CFTs, parametrized by the choice of seed theory $\mathcal{C}$. They have also played a prominent role in the realizations of AdS$_3$/CFT$_2$ that we know from string theory \cite{Maldacena:1997re,Maldacena:1998bw}. 

Unfortunately, it is well known that the light spectrum of symmetric product orbifold theories is universal and exhibits Hagedorn growth \cite{Keller:2011xi}
\be\label{eq:ck1}
\rho(h) \sim e^{2\pi h} \,.
\ee
These theories should in fact be seen as free (discrete) gauge theories. From a holographic standpoint, they are not dual to supergravity in AdS, but rather to string theories in the tensionless limit \cite{Seiberg:1999xz,Eberhardt:2018ouy}. For example, they do not exhibit chaotic dynamics that is characteristic of semi-classical gravity \cite{Belin:2017jli}.

Nevertheless, there could be a connection between symmetric product orbifolds and CFTs that exhibit the supergravity growth we are seeking here. For example, in the known string theory constructions, the symmetric product orbifold is a weakly coupled description of a D-brane configuration wrapping, for example, $\mathbb{T}^4$ or K3; see \cite{Aharony:1999ti,David:2002wn} for a review. In this context the seed CFT is the non-linear sigma model on $\mathbb{T}^4$ or K3, respectively. These CFTs contain a marginal operator which can drive the theory into a strong coupling regime: Under this deformation, most of the states acquire large anomalous dimensions, and the Hagedorn growth \eqref{eq:ck1} is reduced to the much-slower growth \eqref{eq:sugrarho} with $D=6$.

A test of this connection, i.e. that a marginal deformation relates a symmetric product orbifold to a holographic CFT, can be established via two complementary approaches. A first approach is to establish the existence of this marginal operator, and assess that it reduces the growth in \eqref{eq:ck1} to a desired lower growth of the form \eqref{eq:sugrarho}. Recent progress has been made with this regard \cite{Gaberdiel:2015uca,Keller:2019yrr}. This however remains extremely tedious even only at second order in perturbation theory. 

A second approach is to focus on observables that are protected under the marginal deformation. In particular, the spectrum of BPS states in supersymmetric CFTs (with at least ${\cal N}=(2,2)$) is insensitive to this class of deformations. By studying the symmetric product orbifold description, we can therefore test if there is a strong coupling description meeting our criteria \eqref{eq:sugrarho}. We will view this as a necessary condition:  the BPS spectrum will serve as a lamppost to detect gravitational features. 
 
Our main result is a practical implementation of  the second approach,  given only minimal data about the seed theory. The emphasis here will be to quantify the degeneracy of light BPS states of a SCFT in the space of symmetric product orbifold theories. Our analysis gives an exact expression for the degeneracy of states in the large central charge limit, and allows us to compute the spectrum of light operators to any desired precision. We will provide a classification of the seed theories ${\cal C}$ that divides them into two classes:
\begin{description}
\item[\it Hagedorn Landscape:] Supersymmetric examples of symmetric product orbifold theories whose light BPS states have Hagedorn growth as defined above. These instances have less utility in a gravitational context since they will never admit a supergravity growth via a marginal deformation, albeit there might be other interesting CFT or string theory applications of them. This is the most typical situation within the space of symmetric product orbifolds.
\item[\it Supergravity Landscape:] Supersymmetric examples of symmetric product orbifold theories whose light BPS states have supergravity growth. These are promising examples in the sense that a CFT, supergravity or string theory description for the majority of them is unknown to us. Although this category is rather restricted and sparse, we provide a precise landscape of potentially sound instances of AdS$_3$/CFT$_2$ that is for the most part unexplored. 

\end{description}

This work is an improvement of the analysis of \cite{Benjamin:2015vkc}, where the first evidence for these two different behaviors was noticed. This was achieved by studying the specialized  $z=0$  version of the NS elliptic genus. Forms whose specialized genus had supergravity growth were called `very special'. Our analysis shows that the specialized genus is too coarse to distinguish all cases. In particular we will provide an example of a form whose specialized genus has supergravity growth, but whose elliptic genus has Hagedorn growth: specialization in this case leads to additional cancellations.

The paper is organized as follows. We start in Sec.\,\ref{sec:2} by reviewing the properties of the spectrum of two-dimensional CFTs.  We then introduce the landscape of symmetric product orbifolds and describe how their spectrum is obtained from the choice of the seed. We also discuss the spectral flow transformation between Ramond and Neveu-Schwarz sector. In Sec.\,\ref{s:Bad}, we describe the growth of light states for a generic seed and show that it exhibits Hagedorn growth. We then give the precise criterion to distinguish between this case and the situation where the growth is drastically decreased. In Sec.\,\ref{s:Promising}, we give a complete classification of the CFTs that lead to a supergravity-like spectrum and give explicit expressions for the light BPS spectrum. Some of these theories correspond to known string theory constructions, while others are new potential realizations of AdS$_3$/CFT$_2$. We finish in Sec.\,\ref{sec:5} by a discussion of future directions. The detailed analysis of the mathematical techniques to extract the light spectrum is given in a companion paper \cite{Belin:2019jqz}.


\section{Symmetric product orbifolds}\label{sec:2}

In this section, we will review symmetric product orbifolds and describe their universal properties. The emphasis is on their partition functions, and  for supersymmetric instances, their elliptic genera. 


\subsection{Partition functions and elliptic genera in CFT$_2$}
To start, it is instructive to review properties of the density of states for a CFT$_2$ with a discrete spectrum. Let us consider a modular invariant partition function of a chiral CFT with central charge $c$,
\be\label{eq:zc}
Z(\tau)\equiv \Tr_\mathcal{H} q^{L_0-\frac{c}{24}} = \sum_{n\geq-c/24} d(n)q^n~, \qquad q\equiv e^{2\pi i\tau}~.
\ee
It is well known that the asymptotic density of states is given by the Cardy formula, i.e.  $d(n)\sim e^{2\pi \sqrt{cn/6}}$ for states with large $n$ and fixed $c$. The behavior of the heavy states ($n\gg c$) is thus completely determined. The behavior of light states however is not fixed. More precisely, the spectrum of low lying primary states $d(n)$ can be chosen freely for $n \sim O(1)$, in fact roughly up to $n<0$, without tampering with the Cardy formula. A generalization to non-chiral CFTs is straightforward.

Here we will be interested in supersymmetric CFT$_2$, and we will focus mostly on theories with at least $\mathcal{N}=(2,2)$ supersymmetry. For such theories, one can define the elliptic genus as \cite{Witten1987,Eguchi1989,Kawai1994} 
\be \label{ellipticgenus}
\chi(\tau,z)\equiv \Tr_{\scaleto{\text{R,R}}{5pt}} (-1)^{J_0+\bar{J}_0} q^{L_0-\frac{c}{24}} y^{J_0} \bar{q}^{\bar{L}_0-\frac{c}{24}} \,, \quad y\equiv e^{2\pi i z}\,,
\ee
where $z$ is a chemical potential for the $U(1)$ current. Since it generalizes a partition function, $\chi$ has particularly nice transformation properties under the modular group $SL(2,\Z)$. In fact  under modular transformations it behaves like a weak Jacobi form (wJf) of weight 0, meaning it satisfies (\ref{eq:jf1}) as described in appendix~\ref{app:wjf}. If in addition all $U(1)$ charges are integer, it is also invariant under spectral flow, meaning it satisfies (\ref{eq:jf2}). This implies that $\chi(\tau,z)$ is actually a weak Jacobi form of weight 0 and index $t=c/6$. Charge integrality is  ensured for instance if the CFT has $\N=(4,4)$ supersymmetry; or if it is the $\sigma$-model of an even dimensional Calabi-Yau manifold, in which case locality of the holomorphic $\Omega^{d,0}$ form ensures this.

There are however many interesting CFTs for which the elliptic genus has fractional $U(1)$ charges. Typical examples are odd dimensional Calabi-Yau manifolds: the holomorphic $\Omega^{d,0}$ form still ensures that the $U(1)$ charges are integral in the NS sector, but after flowing to the Ramond sector they become half-integer. For CY 3-folds for instance the elliptic genus is a multiple of $\phi_{0,3/2}(\tau,z)$ \cite{Gritsenko:1999fk}, which has half-integer $y$-exponents.
Other examples of ${\cal N}=(2,2)$ with fractional charges are the $\mathbb{T}^4$-orbifolds described in \cite{Datta:2017ert}, or `unorbifolded' Gepner models \cite{Gepner:1987vz}, that is products of $\mathcal{N}=2$ minimal models.
All these elliptic genera do not transform as (\ref{eq:jf2}) anymore.  We can however turn them into a wJf by `unwrapping': Suppose all charges have denominator $k$. Then we define
\be
 \varphi(\tau,z):= \chi(\tau,kz)= \sum_{n,l}c(n,l) q^n y^l
\ee
which does have integer charges, and it is straightforward to check that it indeed defines a wJf of index $t=k^2c/6$. This unwrapping trick thus allows us to also analyze CFTs with fractional charges using wJf. Conversely, given a wJf $\varphi$, its index $t$ may not always correspond to the central charge $c/6$: it could also describe a CFT of smaller central charge with fractional $U(1)$ charges , whose elliptic genus has been unwrapped.
\label{ss:unwrap}

Just like for the partition function, modular invariance gives a Cardy-type formula for the asymptotic behavior of the spectrum of states in $\chi(\tau,z)$. The role of the energy is now effectively played by the discriminant
\be
\Delta=4tn-l^2\ ,
\ee
where $t$ is the index of the weak Jacobi form.
Note that the discriminant is bounded from below by $-t^2$. We will call $\Delta_{\min}$ the discriminant of the state of minimal discriminant.
If the discriminant is large and positive, then the behavior is again given by 
\be\label{eq:susy}
c(n,l) \sim e^{\pi\sqrt{\frac{|\Delta_{\min}|}{t^2}\Delta}}~, \quad \textrm{for}\quad \Delta\gg 1\ .
\ee
(See for instance appendix B of \cite{Belin:2016knb}.) The role of light states is played by states with negative discriminant $\Delta<0$: so-called polar states.  As in \eqref{eq:zc}, their degeneracy does not affect the derivation of \eqref{eq:susy}, and can be chosen (almost) completely freely. The only input in the Cardy-type formula above is the discriminant of the most polar term, $\Delta_{\min}$. 

\subsection{Partition function of symmetric product orbifolds}

In holography, we are interested in families of CFTs with a good large $c$ limit. This means in particular that in the large $c$ limit the CFTs should have a finite number of states at any finite dimension $h$. The best known constructions of this type are symmetric orbifolds, who are part of a larger set of CFTs known as permutation orbifolds \cite{Haehl:2014yla,Belin:2014fna,Belin:2015hwa}. Symmetric orbifolds are constructed in the following way: we start with a seed CFT $\mathcal{C}$ of central charge $c$ and partition function $Z(\tau;\mathcal{C})$, take their $m$-fold tensor product, and then orbifold by all possible permutations of the $m$ factors, that is by the entire symmetric group $S_m$:
\be
\mathcal{C}_m = \mathcal{C}^{\otimes m}/ S_m \,.
\ee
It turns out that the generating function for the partition functions of the $\mathcal{C}_m$ has a very simple form \cite{Dijkgraaf:1996xw}:
\be
\mathcal{Z}(\tau,\rho) = \sum_m Z(\tau; {\cal C}_m)\,p^m = 
\prod_{m>0,n\in\Z}\frac{1}{(1-p^m q^n)^{d(mn)}}=\sum_{m,n} d_m(n)p^mq^n~, \qquad p\equiv e^{2\pi i \rho}~, 
\ee
where $d(n)$ are the degeneracies appearing in $Z(\tau;{\cal C})$.
What can we say about the growth of coefficients $d_m(n)$  of the $m$-th symmetric orbifold? Obviously, for $n$ asymptotically large, that is for $n\gg m$, we will find the universal Cardy behavior \eqref{eq:rho1}. Rather surprisingly however, we also find universal behavior for states with $n<0$. More precisely, regardless of the seed theory we choose, such states have Hagedorn growth \cite{Keller:2011xi},
\be\label{bosHagedorn}
d_m(n) \sim e^{2\pi(n+m\,c/24)}~,\qquad -{m\, c\over24}  \ll n \ll 0\ .
\ee
The idea of the limit here is that we send $m$ to infinity while keeping $n+mc/24$ fixed. The Hagedorn growth in \eqref{bosHagedorn} comes from 
the twisted states: $d_m(n)$ is primarily the sum of contributions of twisted sectors of different length, and its maximal term is already of the form $e^{2\pi (n+mc/24)}$. All other terms are positive, so that the total sum gives the behavior (\ref{bosHagedorn}). The landscape of symmetric orbifold theories thus seems somewhat disappointing: while they all satisfy the HKS sparseness bound \rref{HKS}, the asymptotic behavior of their light states is universally determined, and never corresponds to a holographic dual with a supergravity-like spectrum of perturbative states.

This story becomes much more interesting when we turn to elliptic genera and their symmetric orbifolds. As before, we can write down a very similar looking formula for the generating function \cite{Dijkgraaf:1996xw}:
\be\label{DMVV}
\mathcal{Z}(\tau,z,\rho) = \sum_{m\geq0} \varphi(\tau,z; {\cal C}_m)p^{tm} = 
\prod_{m>0,n,l\in \mathbb{Z}}\frac{1}{(1-p^{tm} q^ny^l)^{c(mn,l)}} = \sum_{m,n,l}d(m,n,l)\,p^{tm}q^n y^l \ .
\ee
Characterizing the growth behavior of the Fourier coefficients $d(m,n,l)$ will be the main object of this work. Note that $d(m,n,l)$ is completely fixed by the choice of the coefficients $c(n,l)$, namely by the choice of a seed theory. Our goal will be to determine, or at least estimate, the coefficients $d(m,n,l)$ given some basic information on the seed.

For heavy states, $\Delta = 4tmn-l^2 \gg 1$, we obtain the universal Cardy behavior \eqref{eq:susy}, as expected. Unlike in the bosonic case discussed above, the behaviour for polar states with $\Delta <0$ can vary drastically! Naively repeating the argument around \eqref{bosHagedorn}, we again find that the maximal twisted sector contributes $e^{2\pi n}$. However, as we are working with an index, there can now be cancellations between terms. Indeed it turns out that for certain examples these cancellations are powerful enough to reduce the Hagedorn growth to a much slower supergravity-like growth. This in fact plays a crucial role in matching the BPS spectrum of CFT light states to gravitational perturbations for the D1-D5 system \cite{deBoer:1998us,deBoer:1998kjm}.

Our goal is to investigate under which circumstances such cancellations happen, and supergravity growth occurs. In order to describe perturbative states, it will be more useful to spectral flow to the NS sector since vacuum AdS is the NS vacuum and perturbative states are those close to the NS vacuum. We therefore describe the spectral flow transformation in the following subsection.

\subsection{NS sector elliptic genus}\label{sec:NS}
Since we are ultimately interested in matching our light states to perturbative states, we want to convert (\ref{DMVV}) to the NS sector. In principle this is a straightforward application of spectral flow. The crucial point however is that to use (\ref{DMVV}), $\varphi$ has to be a wJf. If we want to analyze a CFT with fractional charges, $\varphi$ will therefore be an unwrapping of the actual elliptic genus as discussed in Sec.~\ref{ss:unwrap}, which affects the physical interpretation of what it means to spectrally flow.

Keeping this unwrapping feature in mind, we will identify the NS sector as follows.  Let us denote by $\varphi$ the wJf of weight 0 index $t$ related to the seed  of the symmetric product orbifold. If the most polar term in $\varphi$ is
\be \label{q0yb}
q^0 y^{-b}~,
\ee
we will identify, via spectral flow, this term as the NS vacuum. In general, if the NS vacuum contributes to the NS elliptic genus, it implies that the most polar term is necessarily of the form \rref{q0yb}.\footnote{For the rest of the paper, we will assume that the NS vacuum gives a non-zero contribution to the elliptic genus. In the discussion section, we comment on some aspects of the cases where we relax this assumption.}
The desired transformation is 
\be \label{SF}
\chi_{\scaleto{\text{NS}}{4pt}}(\tau,k\,z;\mathcal{C})=e^{2\pi i \tau \frac{b^2}{4t} }e^{2\pi i t  (\frac{b^2}{4t^2}\tau+\frac{b}{t}z)}\varphi(\tau,z+\frac{b}{2t}\tau) ~,
\ee
where the left hand side is the NS sector elliptic genus unwrapped $k$ times.
This transformation can be viewed in the following way: it is the combination of a spectral flow transformation by  a fractional amount $b/2t$, and an overall shift ---the first exponential term in \eqref{SF}--- that sets the vacuum energy to zero. With this transformation, we guarantee that the NS vacuum is given by the term $q^0y^0$, which will be convenient to take the large $c$ limit. Note that this fixes $bk=t$ and the original Ramond sector elliptic genus has the usual relation to the NS sector, i.e.
\be
\chi_{\scaleto{\text{NS}}{4pt} }(\tau,z) =q^{c\over24} q^{{\hat t}/2} y^{\hat t}\chi (\tau,z +{\tau\over 2})~,
\ee
with $c=6\hat{t}$ the central charge of the seed, and $\hat t= t k^{-2}$ the index of the elliptic genus before unwrapping.

How do we implement such a transformation at the level of the generating function \rref{DMVV}? We simply perform the transformation
\be
t \rho \to t \rho + \frac{b^2}{2t} \tau + b z  \,, \qquad  z \to z+\frac{b}{2t} \tau  \,.
\ee
The generating function becomes
\begin{align}\label{eq:zns1}
\mathcal{Z}_{\scaleto{\text{NS}}{4pt}}&=\prod_{\substack{m>0,\\ n,l\in \mathbb{Z}}} \frac{1}{(1-p^{tm}q^{n+\frac{b}{2t}l+\frac{b^2}{2t}m} y^{l+b m})^{c(n m,l)}}\cr&= \prod_{\substack{m>0,h\geq 0,\\ \lns \in \mathbb{Z}}} \frac{1}{(1-p^{tm}q^h y^{\lns})^{\tilde{c}(m,h,\lns)}} \,,
\end{align}
where $c(n,l)$ are the coefficients in $\varphi$, and we have defined
\bea
h&=& n+\frac{b}{2t}l+\frac{b^2}{2t}m~, \notag \\
\lns&=&l+b m ~,\\
\tilde{c}(m,h,\lns) &=& c(m( h-\frac{b}{2t}\lns),\lns-bm)~. \notag
\eea
Since in the seed $q^0 y^{-b}$ is the most polar term, we can quickly check that the product in \rref{eq:zns1} is indeed only over non-negative values of $h$: Indeed, due to the polarity constraint, $c(m( h-\frac{b}{2t}\lns),\lns-bm)$ is only non-zero if
\be
h \geq \frac{(\lns)^2}{4tm} + \frac{b^2}{4tm}(m^2-1) \geq 0\ .
\ee
The factor in the product formula with $h=0$, $\lns=0$ and $m=1$ corresponds to the vacuum of the seed theory, and reads
\be\label{polep1}
\frac{1}{(1-p^t)^{c(0,-b)}} \,.
\ee
This is the only term with $h=0$ in the product. As we will see below, we will be interested in the behavior of (\ref{eq:zns1}) at $p=1$; (\ref{polep1}) is then the pole in the generating function at that value.

Our aim is to quantify the behavior of light states in $\mathcal{Z}_{\scaleto{\text{NS}}{4pt}}$ in the large central charge limit. Because the vacuum of the seed can be degenerate, i.e. we could have $c(0,-b)>1$, the degeneracy of the vacuum state of the $m$-th copy scales with $m$: This makes the limit $m\to\infty$ of the elliptic genus singular. To regulate this divergence, the procedure to extract the $m\to\infty$ limit is to strip off \eqref{polep1}, and to set $p=1$ in the remaining expression \cite{deBoer:1998us}. Strictly speaking we are thus computing the quantity
\begin{multline}\label{NSprod}
\chi_{\scaleto{\text{NS}}{4pt},\infty} \equiv \lim_{m\to\infty} \frac{\chi_{\scaleto{\text{NS}}{4pt}}(\tau,k z;\mathcal{C}_m)}{m^{c(0,-b)-1}} =\lim_{p\to1}(1-p^t)^{c(0,-b)} \mathcal{Z}_{\scaleto{\text{NS}}{4pt}} \\
= \prod_{\substack{h\geq 0,\lns \in \mathbb{Z}\\ (h,\lns)\neq (0,0)}} \frac{1}{(1-q^hy^\lns)^{\fns(h,\lns)}}
\equiv \sum_{h,\lns}d_\infty^{\text{NS}}(h,\lns)q^h y^{\lns} ~,
\end{multline}
where we defined 
\be\label{eq:fns1}
\fns(h,\lns)= \sum_{m=1}^\infty \tilde{c}(m,h,\lns)\ .
\ee
This object extracts  the spectrum of BPS states with fixed $h$ and $\lns$ in the large central charge limit, i.e. $m\to \infty$. This is the quantity that is best suited to be compared with the perturbative spectrum of a putative dual supergravity theory. 
We will see that there are two outcomes that emerge from the analysis: the coefficients $\fns(h,\lns)$ are either constants, or they exhibit exponential growth. This means the coefficients $d_\infty^{\text{NS}}(h,\lns)$ exhibit either supergravity-like growth \eqref{eq:sugrarho} with $D=2$, or Hagedorn growth.

\section{Hagedorn landscape}\label{s:Bad}

In this section, we will describe the possible outcomes for the coefficients $f_{\NS}(h,\lns)$. We start by giving a general estimate for the growth assuming no cancellations occur and show that we recover the Hagedorn behavior reminiscent of the bosonic partition function. We then give a precise criterion on whether cancellations occur or not, following the work we present in \cite{Belin:2019jqz}. This separates the wJf into two classes: forms with a  Hagedorn growth and forms with a supergravity growth. We conclude this section by giving some details on the regime of validity of the Hagedorn growth in the $(h,\lns)$-plane.

\subsection{The growth of the coefficients $\tf(n,l)$: no cancellations}\label{sec:fnl}

We will now discuss the growth of the coefficients $f_{\NS}(h,\lns)$. 
To this end let us introduce
\be\label{fnl}
\tf(n,l)= \sum_{m=1}^\infty c(n m,l-b m) ~,
\ee
so that 
\be\label{fnsfr}
\fns(h,\lns) = \tf(h-\frac{b}{2t}\lns,\lns)\ ,
\ee
as for practical reasons it is more convenient to work with the $\tf(n,l)$. Note that $n$ must be a non-negative integer which dictates the allowed values of $h$.
As we mentioned in Sec. \ref{sec:NS}, assuming that the NS vacuum contributes to the elliptic genus implies that the most polar term of the wJf $\varphi$ is of the form\footnote{It is worthwhile to mention that for low values of the index, all wJf have this property. However, at high enough index, there exist some finely tuned wJf that are not of this form. We return to this question in the discussion section.}
\be
q^0y^{-b} \,.
\ee

Let us first give a quick argument for the growth of the coefficients that we generically expect. Consider \rref{fnl}, 
and simply estimate the largest term in this sum. Assuming that there are no cancellations, the total sum will have a growth behavior similar to its maximal term. 
For large discriminant, modular invariance gives the following asymptotic behavior of the seed coefficients
\be
c(n,l)\sim \exp \pi \sqrt{4tn-l^2}\frac b t~,
\ee 
which follows from \eqref{eq:susy}. The maximal term in \rref{fnl} occurs for $m=(2tn+bl)/b^2$ and gives
\be
\sim \exp{\pi\sqrt{4n^2+4nlb/t}}\ .
\ee
This implies that
\be\label{fNS}
\fns(h,\lns)\sim \exp 2\pi \sqrt{h^2-\frac{b^2}{4t^2}\lns^2} \,.
\ee
Again, assuming that nothing drastic happens when multiplying out the product (\ref{NSprod}), we expect the same behavior also for the coefficients
\be
d_\infty^{\text{NS}}(h,\lns)\sim \exp 2\pi \sqrt{h^2-\frac{b^2}{4t^2}\lns^2} \,.
\ee
For uncharged states with $\lns=0$ we thus recover Hagedorn growth with slope $2\pi$. This is exactly what we expected from the bosonic analysis: if there are no cancellations, the growth of the elliptic genus should match the growth of the partition function \cite{Keller:2011xi,deLange:2018mri}.

This superficial analysis however misses a very important aspect: Cancellations in fact can occur, and they can lead to a drastically slower growth, as we will now discuss.

\subsection{Growth of the coefficients $\tf(n,l)$: A closer look}

Let us now investigate the growth behavior of the sum \rref{fnl} more carefully.
In the accompanying paper \cite{Belin:2019jqz}, we extract the growth behavior following a detailed analysis that takes into account all possible cancellations. It turns out that one can quickly and efficiently capture the nature of the growth, and in particular separate Hagedorn from supergravity-like growth. The outcome of the analysis is the following: 
First define
\be
f(n,l) = \sum_{m\in\Z} c(nm,l-bm)\ ,
\ee
so that
\be
\tf(n,l)=f(n,l) - c(0,l) - \delta_{n,0}\sum_{m>0} c(0,l+bm)\ .
\ee
Clearly the behavior of $\tf$ and $f$ only differs by very few polar terms, so that we can analyze $f$ instead.

The analysis of \cite{Belin:2019jqz} shows that the growth properties of the $f(n,l)$ can be extracted from a set of specialized versions of the seed wJf  given by $\varphi(\tau,(r\tau+s)/b)$ with $r,s=0,1,\ldots,b$. These specialized forms still have nice modular transformations, albeit under a congruence subgroup of $SL(2,\mathbb{Z})$. Their asymptotic behaviour therefore solely depends on whether there is a negative power of $q$ appearing in the specialized wJf or not. If there is, the $f(n,l)$ exhibit Hagedorn growth. If there is not, they are constants. To determine whether the growth is Hagedorn or not, one simply needs to scan over all the $\varphi(\tau,(r\tau+s)/b)$ and check if we can find a term $q^{-\alpha}$ with $\alpha>0$.

The concrete prescription is the following: for a term $q^ny^{-l}$ of the seed form $\varphi$, define
\be\label{alphacriterion}
\alpha = \max_{j=0,\ldots,b-1}
\left(-\frac{ t}{b^2}j\left(j-\frac{b l}{t}\right)-n\right)\ .
\ee
If $\alpha>0$, this term leads to Hagedorn growth 
\be
f(n,l) \sim \exp 2\pi \sqrt{4\alpha (tn^2/b^2+nl/b)}  \,,
\ee
and consequently
\bea\label{alphaHag}
\fns(h,\lns) &\sim &\exp 2\pi \sqrt{\frac{4t\alpha}{b^2}}\sqrt{h^2-\frac{b^2}{4t^2}\lns} \equiv  \exp \nu \sqrt{h^2-\frac{b^2}{4t^2}\lns} ~.
\eea
Note that if the polarity of the term is positive, then we automatically have $\alpha<0$, which means that a non-polar term cannot produce Hagedorn growth. It is therefore only necessary to test (\ref{alphacriterion}) for the polar terms of the seed, of which there are a finite number at any fixed index. Only if all polar terms lead to $\alpha\leq0$ will we have supergravity growth.

To connect this to the discussion in section~\ref{sec:fnl}, let us first consider the most polar term $y^{-b}q^0$.
In principle, $\alpha$ is maximized for $j=b^2/2t$, for which $\alpha=b^2/4t$ and we get
Hagedorn growth with slope
\be \label{slopemax}
\nu=2\pi\ .
\ee
We see that this is exactly the growth under the non-cancellation assumption that lead to (\ref{fNS}).
The reason why we sometimes have slower growth is that this optimal value of $j$ cannot be attained:
namely it can only be attained if 
\be \label{jinteger}
\frac{b^2}{2t} \in \Z\ .
\ee
If this condition is not satisfied, then we obtain a slower Hagedorn growth or even supergravity growth. 

We can now classify the wJf according to the outcome of evaluating \eqref{alphacriterion}. We will see that the forms leading to supergravity growth are relatively rare, in that at any fixed index, they form only a small subspace of all wJf. It will therefore be easier to give the complete classification by specifying the forms that do have supergravity growth, which will be the object of Sec. \ref{s:Promising}. Before that however, we give some more details on when Hagedorn growth is expected as well as the regime of validity of the Hagedorn growth in the $(h,\lns)$-plane.

\subsection{Forms that lead to Hagedorn growth}

Let us give some more details about the circumstances under which we necessarily have light states that exhibit Hagedorn growth. Consider the most polar term of the wJf
\be
q^0 y^{-b} \,.
\ee
From \rref{alphacriterion}, we
see that if there is a $j=1,\ldots ,b-1$ such that
\be
j-\frac {b^2}t < 0 \ , 
\ee
then we necessarily have $\alpha>0$.
Clearly this is the case if and only if 
\be
b^2 > t\ .
\ee
We are therefore guaranteed to get Hagedorn growth from the most polar term alone if $ b> \sqrt{t}$. This means that there are no supergravity-type forms with $b>\sqrt{t}$. For $b\leq\sqrt{t}$, this particular term does not give Hagedorn growth. We therefore have already found a necessary condition for wJfs to have supergravity growth. It is however not sufficient, since other (less polar) terms could still lead to Hagedorn growth. We will analyze this in detail in section~\ref{s:Promising}.

In Table \ref{t:slope} we summarize the classification of all forms with Hagedorn growth up to index $t=4$. Note that the slope of the Hagedorn growth is indeed not always $2\pi$ due to \rref{jinteger} not being satisfied. Comparing these results with the bosonic partition functions, it means some cancellations have occurred, but not enough to completely kill the Hagedorn behaviour.

The physical consequence of a Hagedorn growth is that it leads to a pole in the free energy at the Hagedorn temperature $T_H$. Interpreting the Hagedorn growth as coming from stringy modes in AdS$_3$, we arrive at the following identity
\be
T_H \equiv \frac{1}{4\pi \ell_s} = \frac{1}{\nu \ell_{\text{AdS}}} \,.
\ee
Therefore, the slope $\nu$ is measuring the string scale in AdS units. Note that the slope is upper bounded by the no-cancellation scenario \rref{slopemax} , which matches with the bosonic partition function \rref{bosHagedorn}. This gives
\be
\frac{\nu}{4\pi}=\frac{ \ell_s }{\ell_{\text{AdS}}}  \leq \frac{1}{2} \,.
\ee
For the examples we find that have $\nu < 2\pi$, it is tempting to associate the non-maximal growth to the presence of an exactly marginal operator that can change the ratio of the string and AdS scales and displaces it from its maximal value. However, unlike the supergravity case, our analysis indicates that there is lower bound for $ \ell_s $, such that the string length cannot be made parametrically small. It would be interesting to understand this phenomenon better and we hope to return to this question in future work.

\begin{table}
	\centering
	\begin{tabular}{|cccc|}
		\hline
		$t$ &  $b$ & $j$ & $\nu$ \\
		\hline
		2 & 2 & 1 & $2\pi$\\
		\hline 
		3 &2 & 1 & $2\pi\sqrt{3/4}$\\
		3 &3 & 1,2 & $2\pi \sqrt{8/9}$\\
		\hline
		4 & 3 & 1 &  $2\pi \sqrt{80/81}$\\
		4 & 4& 2 & $2\pi$\\
		\hline
	\end{tabular}
	\caption{Hagedorn slope for various examples up to $t=4$. $j$ is the integer that maximizes $\alpha$. \label{t:slope}}
\end{table}

Finally let us connect our results to the analysis of \cite{Benjamin:2015vkc}, and in particular to their definition of `very special wJf'.
The example $t=3, b=2$ in Table~\ref{t:slope} is very special under that terminology. This means that the coefficients of the specialized elliptic genus, defined as
\be
\chi_{\text{sp}}(\tau):= \chi_{\infty}^{\text{NS}}(\tau,z=0) \,,
\ee
exhibit supergravity-like growth. Comparing with our results in Table 1, we see here that the full NS elliptic genus has Hagedorn growth with slope $2\pi \sqrt{3/4}$. This means that specializing the elliptic genus to $z=0$ has introduced many additional cancellations and drastically changed its behaviour. This shows that one has to be particularly careful in drawing conclusions from the specialized elliptic genus, since it can introduce further cancellations and therefore be misleading.

\subsection{Hagedorn growth in the $(h,\lns)$-plane} 
Let us now briefly discuss the growth of the $\dinf$ in the Hagedorn case in more detail.
We have concluded that if we find $\alpha>0$,  the $\fns$ have Hagedorn growth (\ref{alphaHag}). We then expect the $\dinf$ to grow similarly, 
\be \label{dinf}
\dinf \sim \exp \nu\sqrt{h^2-\frac{b^2}{4t^2}\lns^2}\ .
\ee
This expression is valid provided we are in the regime
\be
h^2-\frac{b^2}{4t^2}\lns^2 \gg 1 \,.
\ee
In particular, for fixed $\lns$ and large $h$, we recover the usual Hagedorn growth of bosonic partition functions.
We would now like to discuss slightly different regimes. We will summarize our findings in Fig.~\ref{hlnsrays}. 

First, note from (\ref{fnsfr}) that $\fns(h,\lns)=0$ if $h<b\lns/2t$, which gives a type of unitarity bound. If $h$ saturates that bound, that is 
\be
h=\frac{b}{2t}\lns \, ,
\ee
then $\fns(\frac{b}{2t}\lns,\lns )= \tf(0,\lns)$.
From (\ref{fnl}) we then see that $\tf(0,\lns)=0$ for negative $\lns$. For positive $\lns$ there are at most two non-vanishing terms in the sum. This implies that the $\tf(0,\lns)$ are essentially constant,
which leads to the growth 
\be
d_\infty^{\text{NS}}(\frac{b}{2t}\lns,\lns) \sim e^{\sqrt{\lns}} \,.
\ee
It follows that in Fig.~\ref{hlnsrays}, the $\dinf$ vanish below the lower right line, and on the line they have Cardy growth.

Next, consider states which are close to this unitarity bound, namely
\be
\lns=\frac{2t h}{b} -\delta l\,.
\ee
To get $\dinf$, we need to analyze in which ways we can get a state of that particular weight and charge.
To get a state with $h$ and $\lns=2ht/b-\delta l$, we start out with the state $(h-b\delta l /2t, -2ht/b+\delta l)$, and then multiply with the contribution of several states of total weight $b\delta l/2t$ and charge 0. Using \rref{dinf} we get for the multiplicity of such states
\be\label{almostchiralgrowth}
\sim \exp \nu \sqrt{\frac{\delta l^2b^2}{4t^2}} = \exp 2\pi \sqrt{\frac{\alpha}{t}}\delta l\ .
\ee
This means we get Hagedorn growth in $\delta l$.
This is valid as long as $\delta l$ is sufficiently large, but not too large either. More precisely, as soon as $\delta l=2ht/b$, that is as soon as $\lns=0$, this argument breaks down, since the state with which we start would have negative weight. We have indicated this growth by teal arrows in Fig.~
\ref{hlnsrays}.

At this point, this growth will break down.
For $\delta l>0$ fixed and large and growing $h$, we can still trust \rref{dinf} so we find
\be \label{Cardylike}
d_\infty^{\text{NS}}(h,2th/b-\delta l) \sim \exp \nu\sqrt{\frac{b\delta l}{t}h}~,
\ee
which again is sub-exponential.  We find that it matches onto (\ref{almostchiralgrowth}) at $\lns=0$.

Finally consider terms $\fns(h,\lns)$ with negative $\lns$. 
Here there is a similar unitarity bound, which however is more difficult to work out in practice.
When evaluating the term $c(nm,l-bm)$ in (\ref{fnl}), the condition that its discriminant be bigger or equal to the highest polarity is
\be
h\geq \frac{(m^2-1)b^2+\lns^2}{4mt}\ .
\ee
If $\lns< \sqrt{2b}$, then this bound is minimized for $m=1$, giving
\be\label{neglnsboundh}
h\geq \frac{\lns^2}{4t}\ .
\ee
For $\lns \geq \sqrt{2b}$, the bound is minimized for $m= \sqrt{\lns^2-b^2}/b$, and again gives (\ref{neglnsboundh}).
To get the bound for the $d_\infty^{\text{NS}}$, 
we want the smallest ratio for $h/|\lns|$, which  is achieved for $\lns=\pm1$, and given by $h\geq 1/4t$. Note however that because $h=n + b\lns/2t$, this implies that $h$ satisfies at least $h \geq 1/2t$. In total we thus get the bound for the $d_\infty^{\text{NS}}$
\be
 h \geq -\frac\lns{2t}\ .
\ee
Note in fact that if $b=t$, that is if we perform the usual spectral flow, then $h$ is half-integer, so the argument above reduces to the usual unitarity bound $h\geq -\lns/2$.
We summarize the various growth directions in Fig. \ref{hlnsrays}. 

\begin{figure}[H]
	\centering
	{
		\begin{tikzpicture}[scale=1.2]
		
		\begin{axis}[
		unit vector ratio*=1 1 1,
		axis x line=center,
		axis y line=center,xlabel = {$l_{\textrm{NS}}$},
		ylabel = $h$,  yticklabels=\empty, xticklabels=\empty,
		samples=100, every axis x label/.style={
			at={(ticklabel* cs:1)},
			anchor=west, scale=.5
		},
		every axis y label/.style={
			at={(ticklabel* cs:1.0)},
			anchor=south, scale=.5
		}]        
		\addplot [->, name path=A, domain=0:-.8, line width=1, orange] {-x};   
		\addplot [->, name path=A, domain=-.1:-.7, line width=1, orange] {.1-x};   
		\addplot [->, name path=A, domain=-.1:-.52, line width=1, orange] {.28-x};      
		\addplot [->, name path=A, domain=-.1:-.32, line width=1, orange] {.48-x};        
				\addplot [->,name path=B, domain=0:.8, line width=1, magenta] {x};   
		\addplot[->, teal,densely dotted, line width=1]coordinates {(0.3,0.3)(.05,.55)};  
		\addplot[->, teal, densely dotted,line width=1]coordinates {(0.2,0.2)(.05,.35)}; 
		\addplot[->, teal, densely dotted,line width=1]coordinates {(0.1,0.1)(.05,.15)};  		 
		\addplot[->, teal,densely dotted, line width=1]coordinates {(0.4,0.4)(.05,.75)};  
		\end{axis} 
		\end{tikzpicture}
	}
	\caption{The orange solid arrows and the magenta line on the unitarity bound indicate directions of Cardy-like growth \rref{Cardylike}. The torquise  dotted arrows Hagedorn growth \rref{almostchiralgrowth}, which is only valid for $\lns>0$.}
	\label{hlnsrays}
\end{figure}

\section{Supergravity landscape}\label{s:Promising}

In this section, we describe the wJf that lead to a supergravity growth. In principle our methods give a complete classification of all such forms for a given index. 
For concreteness, we give the dimension of the space of such forms up to $t=18$ in Table \ref{t:promising}. We also give explicit expressions of the forms up to $t=9$ in appendix~\ref{app:examples}. We will then review how the known string theory constructions leading to supergravity theories in AdS fall into this class. Such forms we call `good'. Correspondingly, we call `promising' all forms that lead to supergravity growth, but for which we do not know a corresponding string theory construction.

\subsection{Polar terms and weak Jacobi forms}\label{ss:polterms}
Let us now systematically search for wJf which lead to supergravity growth. We demand that the most polar term be of the form $q^0y^{-b}$, but we allow for other terms of the same polarity. From section~\ref{s:Bad} we know that necessarily $b\leq\sqrt{t}$, since otherwise we will automatically have Hagedorn growth. 

In what follows it will be crucial to keep track of the dimension of the space of wJf and their polar terms. Following \cite{Gaberdiel:2008xb}, denote by
\be
j(t) := \dim J_{0,t}
\ee
the dimension of the space of wJf of weight 0 and index $t$, and by 
\be
P(t) := \sum_{k=1}^t \left\lceil \frac{k^2}{4t}\right\rceil
\ee
the number of polar terms in the standard region $0\leq k \leq t$. The central point is then that for $t>4$ there are more polar terms than wJf,
\be
P(t) > j(t) \qquad t>4\ .
\ee
This means that for a choice of polar coefficients, generically there will not be a corresponding wJf. This restricts our ability to construct promising forms.

By using an explicit basis of $J_{0,t}$, we ran a systematic search over all forms up to $t=18$ using (\ref{alphacriterion}) to enforce that there are no terms in the wJf that lead to Hagedorn growth, thus identifying all supergravity growth forms. In Table~\ref{t:promising} we give the dimension of the space of such forms for given $t$ and $b$; dim=0 means that there are no such promising forms, i.e. that a wJf of index $t$ with most polar term $y^{-b}$ does not exist. For index up to $t=9$, we give explicit expressions for the space of promising forms in appendix~\ref{app:examples}.

For $t=1,2,3,4,6$ we find the forms that were already found in \cite{Belin:2018oza}: The seeds are wJf of minimal polarity ($b=1$). For $t=4$ there is in addition the unwrapped form $\phi_{0,1}(\tau,2z)$, where $\phi_{0,1}$ is defined in \eqref{eq:jfex}. Note that special linear combinations of the two promising forms at index four lead to the dihedral orbifolds we will discuss in Sec.\,\ref{dihedral}.  
For $t=5$ the situation becomes more intricate due to the constraints explained above: There is no form for $t=5$ whose only polar term is $y^{-1}$.\footnote{In the process of implementing our search, we noticed that there are no wJfs of minimal polarity ($b=1$) for $7\leq t\leq 18$, which is reflected in Table \ref{t:promising}. This makes the five examples in \cite{Belin:2018oza} look like the exception to a rule.  } However, for $t=5, b=2$ we find a genuinely new form which has polar terms $y^{-2}$ and $y^{-1}$ and whose supergravity coefficients we will provide in the following subsection.

More generally we see that promising forms are relatively rare: the space of all wJf grows like
\be
j(t) \simeq \frac{t^2}{12}\ ,
\ee
and the number of promising wJf grows very slowly.
However, experimentally we find that there is always at least one promising form for every $t, b = \lfloor \sqrt{t} \rfloor$. 

\begin{table}[ht]
	\centering
\begin{tabular}{ccc|}
	$t$ &$b$ &dim 
\\ 1 & 1 & 1
\\ 2 & 1 & 1
\\ 3 & 1 & 1
\\ 4 & 1 & 1
\\ 4 & 2 & 2
\\ 5 & 1 & 0
\\ 5 & 2 & 1
\\ 6 & 1 & 1
\\ 6 & 2 & 2

\\ 7 & 1 & 0

\\ 7 & 2 & 1
\\ & & 

\end{tabular}
\begin{tabular}{ccc|}
$t$ &$b$ &dim 
\\ 8 & 1 & 0

\\ 8 & 2 & 2

\\ 9 & 1 & 0

\\ 9 & 2 & 1

\\ 9 & 3 & 3

\\ 10 & 1 & 0

\\ 10 & 2 & 1

\\ 10 & 3 & 2

\\ 11 & 1 & 0

\\ 11 & 2 & 0

\\ 11 & 3 & 1
\\ & & 

\end{tabular}
\begin{tabular}{ccc|}
$t$ &$b$ &dim 

\\ 12 & 1 & 0

\\ 12 & 2 & 2

\\ 12 & 3 & 3

\\ 13 & 1 & 0

\\ 13 & 2 & 0

\\ 13 & 3 & 1

\\ 14 & 1 & 0

\\ 14 & 2 & 0

\\ 14 & 3 & 1

\\ 15 & 1 & 0

\\ 15 & 2 & 1

\\ 15 & 3 & 2
\end{tabular}
\begin{tabular}{ccc}
$t$ &$b$ &dim 

\\ 16 & 1 & 0

\\ 16 & 2 & 1

\\ 16 & 3 & 2

\\ 16 & 4 & 4

\\ 17 & 1 & 0

\\ 17 & 2 & 0

\\ 17 & 3 & 0

\\ 17 & 4 & 2

\\ 18 & 1 & 0

\\ 18 & 2 & 0

\\ 18 & 3 & 3

\\ 18 & 4 & 3\\

\end{tabular}
\caption{\label{t:promising} Dimension of space of promising forms. When the dimension is zero, it means that a wJf of index $t$ and most polar term $y^{-b}$ does not exist.}
\end{table}



\subsection{Explicit expressions for $f(n,l)$}
In all cases where the growth is supergravity-like, the outcome is that the $\tf(n,l)$ are bounded and essentially constant. In fact for these cases we can give simple closed form expressions for them. To do this, we first remind the definition
\be
\tf(n,l)=f(n,l) - c(0,l) - \delta_{n,0}\sum_{m>0} c(0,l+bm)~.
\ee
The $f(n,l)$ can be computed as described in \cite{Belin:2019jqz}. For promising forms, it turns out that they vanish unless 
\be
n=0 \qquad \textrm{or}\qquad tn+bl=0\ ,
\ee
and that they only depend on $n_b := n \mod b$ and on $l_b := 2(n-n_b)t/b+l \mod b$. 
Note that we automatically have $n_bt/b\in \Z$.
The $f(n,l)$ thus take at most $b^2$ different values. In total we have
\be
f(n,l)= \left\{ \begin{array}{ccl} 
  \sum\limits_{\hat m \in b\Z-l-n_bt/b}c(-n_b \hat m/b-n_b^2t/b^2, \hat m) & : &tn+bl=0
  \ \textrm{or}\  n=0
  \\
  0 & :& \textrm{else}
\end{array}\right.
\ .
\ee
From this it follows that the generating function for the $d(n,l)$ is essentially a ratio of $\theta$-type functions, which is why the $\dinf$ indeed have supergravity growth. In particular, it means we always find a growth\footnote{It is worth mentioning that since we are computing the elliptic genus which counts the difference between bosonic and fermionic operators, we cannot directly deduce the dimension of the internal manifold from our analysis, since there can be further cancellations. This is already apparent in \cite{deBoer:1998us,deBoer:1998kjm} where $D=6$ and the growth is $\exp{h^{1/2}}$; in \cite{Kraus:2006nb}, one has $D=5$ and the growth is $\exp{h^{2/3}}$. }
\be
\rho(h) \sim e^{ h^{1/2}} \,,
\ee
namely a growth of the type \rref{eq:sugrarho} with $D=2$.

As an example for the spectrum of a new promising form, we give the expression for $\chi_\infty^\NS$ for $t=5$ and $b=2$. $\chi_\infty^\NS$  is the $1/5$ spectral flow of the exponential lift of the index 5 form 
\begin{align}\nonumber
\phi_{0,5}^{(2)}(\tau,z)&= \frac{1}{846}\left(\frac{1}{12}\phi_{0,1}^5 + \frac{1}{3}\phi_{0,1}^2\phi_{-2,1}^3E_6+ \frac{1}{6}\phi_{0,1}\phi_{-2,1}^4E_4^2+ \frac{1}{6}\phi_{-2,1}^5E_4E_6+\frac{1}{4}\phi_{0,1}^3\phi_{-2,1}^2E_4\right)\\ 
&= y^{-2}+6y^{-1}+10+6y+y^{2}+\mathcal{O}(q)~,
\end{align}
where the superscript indicates $b$ in $(\ref{t:promising})$. We find
\begin{align}\label{eq:rest5b2}
&\chi_\infty^\NS(\tau,\frac{5}{2}z)=\\ \nonumber
&\prod_{n\geq 1} \frac{(1-q^n)^{10}(1-q^{n-2/5} y^{-2})(1-q^{n-1/5}y^{-1})^6(1-q^{n+1/5} y)^6(1-q^{n+2/5}y^2)}{(1-y^2q^{2/5})^{11}(1-y^{2n+2}q^{(2n+2)/5})^{12}(1-y^{2n-1}q^{(2n-1)/5})^6(1-y^{2n+1}q^{(2n+1)/5})^6}~\\ \nonumber
&\quad  \times \frac{1}{\big(1-q^{(2n-1)}y^{-5(2n-1)}\big)^{12}\big(1-q^{2n}y^{-10n}\big)^{12}}~.
\end{align}

\subsection{Known examples}

In this final subsection, we briefly embed  known realizations of AdS$_3$/CFT$_2$, where a significant portion of the gravitational theory and the CFT are known, into our classification. 

\subsection*{Symmetric products of K3}
The elliptic genus of the sigma model with target space K3 \cite{Kawai1994,Eguchi1989} is given by the following wJf 
\begin{align}
\chi_{\rm K3}(\tau,z) &= 2\phi_{0,1}(\tau,z)\cr 
&= 2y^{-1}+20+2y+2(10y^{-2} -64y^{-1}+108-64y+10y^2)q  +\ldots ~.
\end{align}
In the context of our criteria in Sec.\,\ref{sec:fnl}, the summation in \eqref{eq:fns1} reduces to a sum over Kronecker deltas, and hence does not grow as a function of $(n,l)$. This is the key feature for a supergravity-like growth. More concretely, in the limit $N\to\infty$, the elliptic genus \eqref{NSprod} reads \cite{Aharony:1999ti}
\be
\chi_{\infty}^{\text{NS}}(\tau,z)= \prod_{n=1}^{\infty} \frac{(1-q^n)^{20}(1-q^{n-\frac{1}{2}} y^{-1})^2(1-q^{n-\frac{1}{2}} y)^2}{(1-q^{\frac{n}{2}}y^{n})^{24}(1-q^{\frac{n}{2}}y^{-n})^{24}} \,.
\ee
This expression can be matched by a computation of the 6D ${\cal N}=2$ supergravity spectrum on AdS$_3 \times S^3$ \cite{deBoer:1998us}, which played an important role in the establishment of the duality between string theory on AdS$_3\times S^3 \times$K3 and the symmetric product theory of K3.  Further details of this example can be found in \cite{Belin:2018oza}. Generalizations of this constructions to CHL models are discussed in \cite{JatkarSen2006a,DavidSen2006,PaquetteVolpatoZimet2017}, which correspond to orbifolds acting only on K3.

\subsection*{Orbifolds of $\mathbb{T}^4$ \label{dihedral}}
More recently, a new class of dualities was constructed in \cite{Datta:2017ert}. The dualities relate string theory on AdS$_3\times (S^3\times \mathbb{T}^4)/G$ to the symmetric orbifold theory $(\mathbb{T}^4/G)^{\otimes N}/S_N$. $G$ is a dihedral group and the CFT possesses $\mathcal{N}=(2,2)$ supersymmetry. The dihedral group orbifold is such that the charges are not integer-quantized, but rather half-integer quantized. For this reason, the elliptic genus $\chi(\tau,z)$ is not a true wJf: it is related to one by a simple unwrapping as discussed in Sec.\,\ref{sec:NS}. We have
\begin{align}
\chi^{\gamma}(\tau,2z)=\phi_{0,4}^\gamma(\tau,z)&=\phi_{0,1}(\tau,2z)+\gamma \phi_{0,4}(\tau,z)\cr
&=y^{- 2} + \gamma y^{-1} + 10+\gamma +y^{2} + \gamma y +\mathcal{O}(q) \,.
\end{align}
Here $\phi_{0,4}^\gamma(\tau,z)$ is a wJf of index $t=4$. The value of $\gamma$ depends on the particular dihedral group at hand as well as the choice of discrete torsion and the set of allowed values reads
\be
\gamma=\{ -8,-5,-2,0,2,5,8\} \,.
\ee
Inspecting Table 2, we see that these forms are particular elements of the space of $t=4$ $b=2$ promising forms. One can check that they satisfy the supergravity growth criteria, namely $\alpha<0$ for all polar terms. Supergravity growth was indeed found in \cite{Datta:2017ert} where the elliptic genus $\chi_{\scaleto{\text{NS}}{4pt},\infty}$ is given by 
\bea
\chi_{\scaleto{\text{NS}}{4pt},\infty}(\tau,2z)&=&\prod_{n>0}\left[ \frac{(1-q^n)^{10}(1-q^{n-\frac{1}{2}} y^{-2})(1-q^{n-\frac{1}{2}} y^{2})}{(1-q^{\frac{n}{2}}y^{2n})^{12}(1-q^{\frac{n}{2}}y^{-2n})^{12}} \right] \notag \\
&\times&\left[ \frac{(1-q^n)(1-q^{n-\frac{3}{4}} y)(1-q^{n-\frac{1}{4}} y^{-1})}{(1-q^{\frac{n}{2}}y^{2n})(1-q^{\frac{n}{2}-\frac{1}{4}}y^{2n-1})^2(1-q^{\frac{n}{2}} y^{-2n})^{(-1)^{ n\ \text{mod} \ 2}}} \right]^\gamma \,.
\eea
This function was matched by a supergravity calculation in \cite{Datta:2017ert} which establishes strong evidence for this new class of holographic dualities. 

\section{Discussion}\label{sec:5}

In this paper, we considered the landscape of  two-dimensional SCFTs that are given by symmetric product orbifolds. We studied the elliptic genera of such theories and extracted the growth behavior of the light states. Using a mathematical method described in detail in a companion paper \cite{Belin:2019jqz}, we were able to give the complete classification for the growth of the light states, based on very minimal and concrete data of the seed elliptic genus. The outcome was that there are only two possible types of growth: either the light states exhibit Hagedorn growth or they exhibit supergravity growth of the form \rref{eq:sugrarho} with $D=2$. We now discuss some open questions, in particular how one could uncover new realizations of AdS$_3$/CFT$_2$ from the results presented here.

\subsection{Building new realizations of AdS$_3$/CFT$_2$}

In Sec.\,\ref{s:Promising}, we provided a simple diagnostic to find wJfs whose symmetric product orbifold leads to supergravity growth for the low energy states. We will now give some possible hints towards building the actual CFTs that could correspond to these new promising forms. The first natural step is to consider generalizations of the dihedral orbifolds studied in \cite{Datta:2017ert}. The quotients they analyzed are in principle not the most general orbifolds of $\mathbb{T}^4$ that one can consider. It is thus quite natural to look for higher order symmetry groups of $\mathbb{T}^4$ and orbifold by them, leading to quotients of the form $\mathbb{T}^4/Z_k$. The charges would be fractional and given in units of $1/k$, necessitating an unwrapping to obtain a wJf. Therefore, the index of the wJf would be $t=k^2$ with $b=k$.  Comparing with the entries in Table 2, we indeed find supergravity growth for $t=9,b=3$ and $t=16,b=4$ which sounds very promising. On the gravity side, one would be looking for backgrounds of the form AdS$_3\times (S^3 \times \mathbb{T}^4)/Z_k$. The question is simply to check if one can combine the quotient on the sphere and the one on the torus in such a way to preserve supersymmetry. It would be very interesting to check this explicitly and we hope to return to this question in the future.

More generally, one can also consider other CFTs leading to ${\cal N}=2$ supersymmetry with fractional charge. One interesting route is to consider products of supersymmetric minimal models. We expect certain combinations to lead to promising wJfs, after a suitable unwrapping. It would be very interesting to see if most promising forms can be recovered this way or not. We leave this question for future work.

From a supergravity perspective, there have been several interesting developments that could connect with our approach. In particular it is worth considering, for example,  solutions in type IIB supergravity \cite{Kim:2005ez,Gauntlett:2006af},  the recent F-theory constructions in \cite{Couzens:2017nnr,CouzensLawrieMartelliEtAl2017}, and the backgrounds of massive IIA developed in \cite{Macpherson:2018mif,Lozano:2019emq,Lozano:2019jza,Lozano:2019zvg,Lozano:2019ywa}. In addition, the authors in \cite{ArabiArdehali:2018mil,Couzens:2019wls} study the perturbative supergravity spectrum and quantum effects on the central charge: both of these could be compared with our results if there is a suitable match.

\subsection{Elliptic genus without a NS vacuum}

In the bulk of this paper, we have assumed that the NS elliptic genus receives a non-zero contribution from the (possibly degenerate) vacuum. One may ask what happens if we relax this assumption. In such a scenario the seed has a NS elliptic genus given by
\be
\chi^{\NS}=q^{h_{\min}} y^{l_{\min}} \,,
\ee
for some $h_{\min}>0$ and $l_{\min}$ that we leave unspecified. Once we take the symmetric product, the gap between the (vanishing) vacuum and the first excited state increases, and for the $m$-th orbifold it is $m h_{min}$. This means that there are no light states per se, as all non-vanishing states have a dimension that scales with $m$. Nevertheless, one could perform an additional shift and consider the growth of states close to this lightest non-vanishing state.

Interestingly, once one flows to the Ramond sector this is connected to studying more general types of wJf, in particular it can happen that the most polar term is not of the form $q^0y^{-b}$, but rather has some positive power of $q$. It would be interesting to understand the growth for such a choice of wJf. Unfortunately, we are currently unable to extract the growth of states for such forms using the technology of \cite{Belin:2019jqz} but one should be able to generalize our method to include these types of forms.

\section*{Acknowledgements}
We are happy to thank Vassilis Anagiannis, Nathan Benjamin, Jan de Boer, Lorenz Eberhardt, Matthias Gaberdiel, Robert Maier and Natalie Paquette for useful discussions. AB is partly supported by the NWO VENI grant 680-47-464 / 4114. AC is supported by Nederlandse Organisatie voor Wetenschappelijk Onderzoek (NWO) via a Vidi grant.  This work is supported by the Delta ITP consortium, a program of the Netherlands Organisation for Scientific Research (NWO) that is funded by the Dutch Ministry of Education, Culture and Science (OCW). The work of BM is part of the research programme of the Foundation for Fundamental Research on Matter (FOM), which is financially supported by the Netherlands Organisation for Science Research (NWO).

\appendix

\section{Weak Jacobi Forms}\label{app:wjf}


A weak Jacobi form $\varphi_{k,m}(\tau,z)$ \cite{MR781735} is a holomorphic function on $\H\times\CC\rightarrow\CC$ that has a Fourier expansion
\be\label{eq:jf3}
\varphi_{k,m}(\tau,z)= \sum_{\substack{n\geq0,l}} c(n,l) q^n y^l~,\qquad q = e^{2\pi i\tau}\ , \qquad y = e^{2\pi i z}~,
\ee
and satisfies the transformation properties 
\be\label{eq:jf1}
\varphi_{k,m}\le({a\tau+b\over c\tau +d},{z\over c\tau +d}\ri)= (c\tau +d)^k\exp\le({2\pi i m c z^2\over c\tau +d}\ri)\varphi_{k,m}(\tau,z)~,\quad  \forall \twobytwo{a}{b}{c}{d} \in SL(2,\ZZ)~,
\ee
and 
\be\label{eq:jf2}
\varphi_{k,m}\le(\tau,{z+ \lambda \tau +\mu}\ri)= \exp\le(-{2\pi i m (\lambda^2\tau+2\lambda z +\mu)}\ri)\varphi_{k,m}(\tau,z)~, \quad \lambda,\mu \in \ZZ~.
\ee
Here $k$ is the \emph{weight} and $m$ is the \emph{index} of $\varphi_{k,m}(\tau,z)$.  We define the \emph{discriminant} of a term as $\Delta:= 4nm-l^2$.
The coefficients $c(n,l)$ then only depend on $\Delta$ and $l$ (mod $2m$),
and in fact only on $\Delta$ if $m$ is prime. 
%
%
Moreover one can show that we have $c(n,l)=0$ if
$\Delta < - m^2$, leading to a Fourier expansion
\be
\varphi_{k,m}(\tau,z)= \sum_{\substack{n\geq0,l\\ 4mn- l^2\geq -m^2}} c(n,l) q^n y^l\ .
\ee

The ring of even weight weak Jacobi forms is freely generated by the forms
\bea
E_4(\tau) &=& 1 + 240 \sum_{n=1}^\infty \sigma_3(n) q^n~,\\
E_6(\tau) &=& 1 - 504 \sum_{n=1}^\infty \sigma_5(n) q^n~,\\
\phi_{0,1}(\tau,z) &=& 4\left(\frac{\theta_2(\tau,z)^2}{\theta_2(\tau,0)^2} + \frac{\theta_3(\tau,z)^2}{\theta_3(\tau,0)^2} + \frac{\theta_4(\tau,z)^2}{\theta_4(\tau,0)^2}\right)~, \label{eq:jfex}\\
\phi_{-2,1}(\tau,z) &=& -\frac{\theta_1(\tau,z)^2}{\eta(\tau)^6}~.
\eea

\section{Weak Jacobi Forms with supergravity growth}\label{app:examples}

Below we list the promising wJf of table $(\ref{t:promising})$ up to index $t=9$. Here the superscript on the Jacobi forms indicates $b$ in $(\ref{t:promising})$. 

\begin{align}
\nonumber
\phi_{0,1}^{(1)}(\tau,z)&= \alpha\phi_{0,1}=\alpha\left(y^{-1}+10+y\right)+\mathcal{O}(q)~,\\ \nonumber
\quad \phi_{0,2}^{(1)}(\tau,z)&= \alpha\left(\phi_{0,1}^2 -\phi_{-2,1}^2E_4\right)=24\alpha\left(y^{-1}+4+y\right)+\mathcal{O}(q)~,\\ \nonumber
 \quad \phi_{0,3}^{(1)}(\tau,z)&= \alpha\left(\phi_{0,1}^3-3\phi_{-2,1}^2\phi_{0,1}E_4+2\phi_{-2,1}^3E_6\right)=432\alpha\left(y^{-1}+2+y\right)+\mathcal{O}(q)~,\\ \nonumber
\phi_{0,4}^{(1)}(\tau,z)&= \alpha\Big(\phi_{0,1}^4-6\phi_{-2,1}^2\phi_{0,1}^2E_4+8\phi_{-2,1}^3\phi_{0,1}E_6-3\phi_{-2,1}^4E_4^2\Big)~\\ \nonumber
&=6912\alpha\left(y^{-1}+1+y\right)+\mathcal{O}(q)~,\\ \nonumber
\phi_{0,4}^{(2)}(\tau,z)&=\alpha\phi_{0,1}^4+\beta\phi_{-2,1}^3\phi_{0,1}E_6 -(2\alpha+\beta/2)\phi_{-2,1}^2\phi_{0,1}^2E_4+(\alpha-\beta/2)\phi_{-2,1}^4E_4^2~\\ \nonumber
&=72\left((8\alpha-\beta)y^{-2}+4(16\alpha+\beta)y^{-1}+6(24\alpha-\beta)+\cdots\right)+\mathcal{O}(q)~,\\ \nonumber
\phi_{0,5}^{(2)}(\tau,z)&= \alpha\left(\phi_{0,1}^5 + 3\phi_{-2,1}^4\phi_{0,1}E_4^2+2\phi_{-2,1}^3\phi_{0,1}^2E_6-4\phi_{-2,1}^2\phi_{0,1}^3E_4-2\phi_{-2,1}^5E_4E_6\right) ~\\ \nonumber
&=10368\alpha\left(y^{-2}+6y^{-1}+10+\left(y^{-1}\leftrightarrow y\right)\right)+\mathcal{O}(q)~,\\ \nonumber
\phi_{0,6}^{(1)}(\tau,z)&=\alpha\Big(\phi_{0,1}^6+24\phi_{-2,1}^5\phi_{0,1}E_4 E_6 -45\phi_{-2,1}^4\phi_{0,1}^2 E_4^2
+40\phi_{-2,1}^3\phi_{0,1}^3E_6-15\phi_{-2,1}^2\phi_{0,1}^4 E_4\\ \nonumber
&-32\phi_{-2,1}^6E_6^2+ 27\phi_{-2,1}^6 E_4^3 \Big)\cr
&= 1492992 \alpha\left(y^{-1}+y\right)+\mathcal{O}(q)~,\\ \nonumber
\phi_{0,6}^{(2)}(\tau,z)&=\alpha\left(\phi_{0,1}^6-36\phi_{-2,1}^5\phi_{0,1}E_4 E_6+45\phi_{2,1}^4\phi_{0,1}^2 E_4^2-20\phi_{-2,1}^3\phi_{0,1}^3E_6+28\phi_{-2,1}^6E_6^2-18\phi_{-2,1}^6 E_4^3 \right)\\ \nonumber
&+\beta\Big(6\phi_{2,1}^4\phi_{0,1}^2 E_4^2-4\phi_{-2,1}^5\phi_{0,1}E_4 E_6- 4\phi_{-2,1}^3\phi_{0,1}^3E_6+\phi_{-2,1}^2\phi_{0,1}^4 E_4+4\phi_{-2,1}^6E_6^2-3\phi_{-2,1}^6 E_4^3  \Big)~\\ \nonumber
&=20736\left((15\alpha+\beta)y^{-2}+4(3\alpha-\beta)y^{-1}+6(15\alpha+\beta)+\left(y^{-1}\leftrightarrow y\right)\right)+\mathcal{O}(q)~,\\ \nonumber
\phi_{0,7}^{(2)}(\tau,z)&= \alpha\Big(\phi_{0,1}^7+16\phi_{-2,1}^6\phi_{0,1}E_6^2+9\phi_{-2,1}^6\phi_{0,1}E_4^3-36\phi_{-2,1}\phi_{0,1}^2E_4E_6+15\phi_{-2,1}^4\phi_{0,1}^3E_4^2\\ \nonumber
&+10\phi_{-2,1}^3\phi_{0,4}^4E_6-9\phi_{-2,1}^2\phi_{0,1}^5E_4-6\phi_{-2,1}^7E_4^2E_6\Big)~\\ \nonumber
&=2985984\alpha\left(y^{-2}+3y^{-1}+4+\left(y^{-1}\leftrightarrow y\right) \right)+\mathcal{O}(q)~,\\ \nonumber
\phi_{0,8}^{(2)}(\tau,z)&= \alpha\Big(\phi_{0,8}^2-1440\phi_{-2,1}^7\phi_{0,1}E_4^2E_6+352\phi_{-2,1}^6\phi_{0,1}^2E_6^2- 336\phi_{-2,1}^5\phi_{0,1}^3E_4E_6\\ \nonumber
&+210\phi_{-2,1}^4\phi_{0,1}^4E_4^2-56\phi_{-2,1}^3\phi_{0,1}^5E_6-72\phi_{-2,1}^6\phi_{0,1}^2E_4^3+117\phi_{-2,1}^8E_4^4-96\phi_{-2,1}^8E_4E_6^2 \Big)~\\ \nonumber
&+\beta\Big(\phi_{-2,1}^2\phi_{0,1}^6E_4-20\phi_{-2,1}^7\phi_{0,1}E_4^2E_6+24\phi_{-2,1}^6\phi_{0,1}^2E_6^2- 20\phi_{-2,1}^5\phi_{0,1}^3E_4E_6+15\phi_{-2,1}^4\phi_{0,1}^4E_4^2 \\ \nonumber
&-6\phi_{-2,1}^3\phi_{0,1}^5E_6-9\phi_{-2,1}^6\phi_{0,1}^2E_4^3+9\phi_{-2,1}^8E_4^4-8\phi_{-2,1}^8E_4E_6^2\Big) ~\\ \nonumber
&=2985984\left((28\alpha+\beta)y^{-2}-4(4\alpha+\beta)y^{-1}+6(20\alpha+\beta)+\left(y^{-1}\leftrightarrow y\right) \right)+\mathcal{O}(q)~\\ 
\phi_{0,9}^{(2)}(\tau,z)&=\alpha\Big(\phi_{0,1}^9+144\phi_{-2,1}^8\phi_{0,1}E_4E_6^2-82\phi_{-2,1}^8\phi_{0,1}E_4^4-23328\phi_{-2,1}^7\phi_{0,1}^2E_4^2E_6\\ \nonumber
&+162\phi_{-2,1}^6\phi_{0,1}^3E_4^3+48\phi_{-2,1}^6\phi_{0,1}^3E_6^2-126\phi_{-2,1}^5\phi_{0,1}^4E_4E_6-18\phi_{-2,1}^2\phi_{0,1}^6E_4\\ \nonumber
&+42\phi_{-2,1}^3\phi_{0,1}^6E_6+54\phi_{-2,1}^9E_4^3E_6-64\phi_{-2,1}^9E_6^3\Big)~\\ \nonumber
&=644972544\alpha\left(y^{-2}+2y^{-1}+2+\left(y^{-1}\leftrightarrow y\right)\right)+\mathcal{O}(q)~,\\ \nonumber
\phi_{0,9}^{(3)}(\tau,z)&=\alpha\Big(\phi_{0,1}^9-459 \phi_{-2,1}^8\phi_{0,1}E_4^4+144\phi_{-2,1}^8\phi_{0,1}E_4E_6^2+ 103680\phi_{-2,1}^7\phi_{0,1}^2E_4^2E_6-624\phi_{-2,1}^6\phi_{0,1}^3E_6^2\\ \nonumber
&-216\phi_{-2,1}^6\phi_{0,1}^3E_4^3+504\phi_{-2,1}^5\phi_{0,1}^4E_4E_6-123\phi_{-2,1}^4\phi_{0,1}^5E_4^2+216\phi_{-2,1}^9E_4^3E_6-80\phi_{-2,1}^9E_6^3\Big)~\\ \nonumber
&+\beta\Big(\phi_{-2,1}^2\phi_{0,1}^6E_4
-63\phi_{-2,1}^8\phi_{0,1}E_4^4+28\phi_{-2,1}^8\phi_{0,1}E_4E_6^2+ 12096\phi_{-2,1}^7\phi_{0,1}^2E_4^2E_6\\ \nonumber
&-84\phi_{-2,1}^6\phi_{0,1}^3E_6^2-21\phi_{-2,1}^6\phi_{0,1}^3E_4^3+70\phi_{-2,1}^5\phi_{0,1}^4E_4E_6-21\phi_{-2,1}^4\phi_{0,1}^5E_4^2+30\phi_{-2,1}^9E_4^3E_6\\ \nonumber
&-12\phi_{-2,1}^9E_6^3~
\Big)+\gamma\Big( \phi_{-2,1}^3\phi_{0,1}^6E_6+12\phi_{-2,1}^8\phi_{0,1}E_4E_6^2+2160\phi_{-2,1}^7\phi_{0,1}^2E_4^2E_6\\ \nonumber
&- 20\phi_{-2,1}^6\phi_{0,1}^3E_6^2+15\phi_{-2,1}^5\phi_{0,1}^4E_4E_6-18\phi_{-2,1}^8\phi_{0,1}E_4^4-6\phi_{-2,1}^4\phi_{0,1}^5E_4^2+9\phi_{-2,1}^9E_4^3E_6\\ \nonumber
&-4\phi_{-2,1}^9E_6^3\Big)\cr
&=248832\Big((84\alpha+7\beta+\gamma)y^{-3}-6(12\alpha+5\beta+\gamma)y^{-2}+3(276\alpha+19\beta+5\gamma)y^{-1}\\ \nonumber
&+4(12\alpha-17\beta-5\gamma)+\left(y^{-1}\leftrightarrow y\right)\Big)+\mathcal{O}(q)~,
\end{align}

\bibliographystyle{JHEP-2}
\bibliography{ref}

\end{document}